\documentclass[12pt]{amsart}
 \theoremstyle{definition}
 
 \theoremstyle{remark}

 \numberwithin{equation}{section}

\begin{document}
\title[automorphisms of quantum probabilistic maps]
{Characterization of affine automorphisms and ortho-order
automorphisms of quantum probabilistic maps}
\author{Zhaofang Bai}
\address[Zhaofang Bai]{School of Mathematical Sciences, Xiamen University,
Xiamen, 361000, P. R. China.}
\author{Shuanping Du$^*$}
\address[Shuanping Du]{School of Mathematical Sciences, Xiamen University,
Xiamen, 361000, P. R. China.} \email[Shuanping
Du]{shuanpingdu@yahoo.com}
\thanks{$^*$  Corresponding author}
%\thanks{This paper is in final form and no version of it will be
%submitted for publication elsewhere.}

\maketitle

\begin{abstract}
In quantum mechanics, often it is important for the representation
of quantum system to study the structure-preserving bijective maps
of the quantum system. Such maps are also called isomorphisms or
automorphisms. In this note, using the Uhlhorn-type of Wigner's
theorem, we characterize all affine automorphisms and ortho-order
automorphisms of quantum probabilistic maps.

\end{abstract}

\maketitle

\section{Introduction }

In quantum physics, of particular importance for the
representation of physical system and symmetries are
structure-preserving bijective maps of the system. Such maps are
also called isomorphisms or automorphisms. Automorphisms or
isomorphisms are frequently amenable to mathematical formulation
and can be exploited to simplify many physical problems. By now,
they have been extensively studied in different quantum systems,
and systematic theories have been achieved \cite{Lud}. Recently,
most deepest results in this field have been obtained by Lajos
Monlar in a series of articles
\cite{Mol2,Mol3,Mol4,Mol5,Mol6,Mol7}. And an overview of recent
results can be found in \cite{Cass2,Mol}.

Let us now fix the notations and set the problem in mathematical
terms. Let $H$ be a separable Hilbert space with dimension at
least 3 and inner product $<.>$. Let ${\mathcal B_1}(H)$ be the
complex Banach space of the trace class operators on $H$, with
trace $tr(T)$ and trace norm $\|T\|_1=tr(|T|)$, $|T|=\sqrt{T^*T}$,
$T\in{\mathcal B_1}(H)$. The self-adjoint part of ${\mathcal
B_1}(H)$ is denoted by ${\mathcal B_{1r}}(H)$ which is a real
Banach space. By ${\mathcal B_{1r}}^+(H)$ we denote  the positive
cone of ${\mathcal B_{1r}}(H)$. As usual, the unit ball of
${\mathcal B_{1r}}^+(H)$ is denoted by $S_1(H)=\{T\in{\mathcal
B_{1r}}^+(H):tr(T)=\|T\|_1\leq 1\}$, the surface of $S_1(H)$ by
$V=\{T\in{\mathcal B_{1r}}^+(H):tr(T)=1\}$. With reference to the
quantum physical applications, ${\mathcal B_{1r}}(H)$ is called
state space, the elements of ${\mathcal B_{1r}}^+(H)$ and  $V$ are
called density operators and states, respectively (see \cite{Kad},
\cite{Lud}). Naturally,  $S_1(H)$ can be equipped with several
algebraic operations. Clearly, $S_1(H)$ is a convex set, so one
can consider the convex combinations on it. Furthermore, one can
define a partial addition on it. Namely, if $T,S\in S_1(H)$ and
$T+S\in S_1(H)$, then one can set $T\oplus S=T+S$. Moreover, as
for a multiplicative operation on $S_1(H)$, note that in general,
$T,S\in S_1(H)$ does not imply that $TS\in S_1(H)$. However, we
all the time have $TST\in S_1(H)$, since $TST\in {\mathcal
B_{1r}}^+(H)$ and $tr(TST)=\|TST\|_1\leq\|T\|_1\|S\|_1\|T\|_1\leq
1$. This multiplication  is a nonassociative operation and
sometimes called Jordan triple product also appears in infinite
dimensional holomorphy as well as in connection with the
geometrical properties of $C^*$-algebras. Finally, there is a
natural partial order $\leq$ on $S_1(H)$ which is induced by the
usual order between selfadjoint operators on $H$. So, for any
$T,S\in S_1(H)$ we write $T\leq S$ if and only if $<Tx,x> \leq
<Sx,x>$ holds for every $x\in H$. Physically, the most interesting
order may be spectral order (see \cite{Ols}). The detailed
definition is as follows. Let $T,S\in S_1(H)$ and consider their
spectral measures $E_T,E_S$ defined on the Borel subsets of
${\mathbb R}$. We write $$T\preceq S \mbox{\hspace{0.1in}if and
only if \hspace{0.1in}} E_T(-\infty,t]\geq
E_S(-\infty,t]\hspace{0.1in}(t\in{\mathbb R}).$$ The spectral
order has a natural interpretation in quantum mechanics. In fact,
the spectral projection $E_T(-\infty,t]$ represents the
probability that a measurement of  $T$ detects its value in the
interval $(-\infty,t]$. Hence for $T,S\in S_1(H)$ the relation
$T\preceq S$ means for every $t\in[0,1]$ we have
$E_T(-\infty,t]\geq E_S(-\infty,t]$ in each state of the system,
i.e., the corresponding distribution functions are pointwise
ordered.

Because of the importance of $S_1(H)$, it is a natural problem to
study the automorphisms of the mentioned structures. The aim of
this paper is to contribute to these investigations. In \cite{BD},
the automorphisms of $S_1(H)$ with the partial addition and Jordan
triple product were characterized. In this paper, we are aimed to
characterize the affine automorphisms and ortho-order
automorphisms of $S_1(H)$. The core of the proof is to reduce the
problem to using the Uhlhorn-type of Wigner's theorem (see
\cite{Uh}).

Now, let us give the concrete definitions of affine automorphism
and ortho-order automorphism. A bijective map
$\Phi:S_1(H)\rightarrow S_1(H)$ is an affine automorphism if
$$\Phi(\lambda T+(1-\lambda)S)=\lambda\Phi(T)+(1-\lambda)\Phi(S) \hspace{0.1in}\mbox{for all} \hspace{0.1in}T,S\in S_1(H),0\leq\lambda\leq 1.$$
A bijective map $\Phi:S_1(H)\rightarrow S_1(H)$ is called an
ortho-order automorphism if
$$\begin{aligned}{\rm (i)}&\hspace{0.1in} TS=0\Leftrightarrow\Phi(T)\Phi(S)=0
\hspace{0.1in}\mbox{for all}\hspace{0.1in} T,S\in S_1(H),\\
{\rm(ii)}&\hspace{0.1in}T\preceq
S\Leftrightarrow\Phi(T)\preceq\Phi(S)\hspace{0.1in}\mbox{for
all}\hspace{0.1in} T,S\in S_1(H).\end{aligned}$$

Here, it is worth mentioning that the affine automorphism has an
intimate relationship with the so-called operation of ${\mathcal
B}_1(H)$(see \cite{HR,Kra}), which is a fundamental notion in
quantum theory. Recall that an operation $\Phi$ is a completely
positive linear mapping on ${\mathcal B}_1(H)$ such that $0\leq
tr(\Phi(T))\leq 1$ for every $T\in V$. An operation represents a
probabilistic state transformation. Namely, if $\Phi$ is applied
on an input state $T$, then the state transformation
$T\rightarrow\Phi(T)$ occurs with the probability $tr(\Phi(T))$,
in which case the output state is $\frac{\Phi(T)}{tr(\Phi(T))}$.
By the Kraus representation theorem (see \cite{BP,Kra}), $\Phi$ is
an operation if and only if there exists a countable set of
bounded linear operators $\{A_k\}$ such that $\sum_{k}
A_k^*A_k\leq I$ and $\Phi(T)=\sum_{k}A_kTA_k^*$ holds for all
$T\in{\mathcal B}_1(H)$. This is very important in describing
dynamics, measurements, quantum channels, quantum interactions,
quantum error, correcting codes, etc \cite{NC}. Since operation
$\Phi$ is linear and $0\leq tr(\Phi(T))\leq 1$ for every $T\in V$,
it is evident such $\Phi$ maps $S_1(H)$ into $S_1(H)$ and
possesses the affine condition mentioned in the definition of
affine automorphism. Thus operations on ${\mathcal B}_1(H)$ can be
reduced to maps on $S_1(H)$. Furthermore, if the reduction of
operation on $S_1(H)$ is bijective,  from our Theorem 2.1, an
explicit description can be given even without completely positive
assumption.

\section{Affine automorphisms on $S_1(H)$}

In this section, we present a structure theorem of affine
automorphisms on $S_1(H)$. The following are the main results of
this section.

\textbf{Theorem 2.1.} {\it If $\Phi:S_1(H)\rightarrow S_1(H)$ is
an affine automorphism, then there exists an either unitary or
antiunitary operator $U$ on $H$ such that $\Phi(T)=UTU^*$ for all
$T\in S_1(H)$.}

\textbf{Corollary 2.2.} {\it If $\dim H<+\infty$, $\Phi$ is a
$\|.\|_1$-isometric automorphism of $S_1(H)$ (which is a bijection
of $S_1(H)$ and satisfies $\|T-S\|_1=\|\Phi(T)-\Phi(S)\|_1$ for
all $T,S\in S_1(H)$), then there exists an either unitary or
antiunitary operator $U$ on $H$ such that $\Phi(T)=UTU^*$ for all
$T\in S_1(H)$.}

We remark that, in the above two results, the bijectivity
assumption is indispensable to obtain a nice form of $\Phi$. To
show it, an example originating from the Kraus representation
theorem will be given after the proof of theorem 2. 1 and
corollary 2.2.

Before the proof of Theorem 2.1, let us recall the general
structure of density operators (see for instance \cite{BB}). For
$T\in{\mathcal B}_{1r}^+(H)$, there exists an orthonormal basis
$\{e_n\}_{n\in{\mathbb N}}$ of $H$ and numbers $\lambda_n>0$ such
that $$T=\sum_{n=1}^{+\infty }\lambda_n P_n$$ or
$$Tx=\sum_{n=1}^{+\infty}\lambda_n<x,e_n>e_n, \forall
x\in H\hspace{0.1in}\mbox{and}\hspace{0.1in}
0<tr(T)=\sum_{n=1}^{+\infty}\lambda_n<+\infty,$$  where $P_n$ is
the one dimensional projection onto the eigenspace spanned by the
eigenvector $e_n$. Let ${\mathcal P}_1(H)$ stand for the set of
all one dimensional projections on $H$.  With reference to the
quantum physical applications, the elements of ${\mathcal P}_1(H)$
are called pure states.

Now, we are in a position to prove our first theorem.

{\bf Proof of Theorem 2.1.}  we will finish the proof by checking
3 claims.

{\bf Claim 1.} $\Phi$ is continuous in the trace norm.

To see the continuity of $\Phi$, consider the affine
transformation $\Psi:S_1(H)\longmapsto{\mathcal B_{1r}}(H)$
defined by $\Psi(T)=\Phi(T)-\Phi(0)$ for every $T\in S_1(H)$. It
is easy to see that $\Psi$ is injective. In the following, we will
prove $\Psi$ has a unique linear  extension from $S_1(H)$ to
${\mathcal B}_{1r}(H)$. Since $\Psi$ is affine and $\Psi(0)=0$,
for each $\lambda\in[0,1]$ and every $T\in S_1(H)$, $\Psi(\lambda
T)=\lambda\Psi(T)$. For $T\in{\mathcal B}_{1r}^+(H)$, a natural
extension of $\Psi$ from $S_1(H)$ to ${\mathcal B}_{1r}^+(H)$ is
to define
$$\widetilde{\Psi}(T)=\|T\|_1\Psi(\frac{T}{\|T\|_1}).$$
Then for any $\lambda \geq 0$, one gets $\widetilde{\Psi}(\lambda
T)=\lambda\widetilde{\Psi}(T)$, which is the positive homogeneity.
For $T,S\in{\mathcal B}_{1r}^+(H)$, suppose
$\widetilde{\Psi}(T)=\widetilde{\Psi}(S)$, without loss of
generality, assume further $\|T\|_1\leq\|S\|_1$. Then
$\widetilde{\Psi}(\frac{T}{\|S\|_1})=\widetilde{\Psi}(\frac{S}{\|S\|_1}),\frac{T}{\|S\|_1},\frac{S}{\|S\|_1}\in
S_1(H)$. By the injectivity of $\Psi$, $T=S$ and thus $
\widetilde{\Psi}$ is injective. For $T_1,T_2\in{\mathcal
B}_{1r}^+(H)$, we can rewrite $T_1+T_2$ in the form
$$T_1+T_2=(\|T_1\|_1+\|T_2\|_1)(\frac{\|T_1\|_1}{\|T_1\|_1+\|T_2\|_1}\frac{T_1}{\|T_1\|_1}+
\frac{\|T_2\|_1}{\|T_1\|_1+\|T_2\|_1}\frac{T_2}{\|T_2\|_1}).$$ The
positive homogeneity of  $\widetilde{\Psi}$ and the affine
property of $\Psi$ yield the additivity of $\widetilde{\Psi}$,
that is,
$\widetilde{\Psi}(T_1+T_2)=\widetilde{\Psi}(T_1)+\widetilde{\Psi}(T_2)$.

Next for $T\in{\mathcal B}_{1r}(H)$, write $T=T^+-T^-$, where
$T^+=\frac{1}{2}(|T|+T), T^-=\frac{1}{2}(|T|-T),
|T|=(T^*T)^\frac{1}{2}$. Let
$$\widehat{\Psi}(T)=\widetilde{\Psi}(T^+)-\widetilde{\Psi}(T^-).
$$
Also if $T=T_1-T_2$ for some other $T_1,T_2\in{\mathcal
B}_{1r}^+(H)$, then $T^++T_2=T^-+T_1$, by the additivity of
$\widetilde{\Psi}$,
$\widetilde{\Psi}(T^+)-\widetilde{\Psi}(T^-)=\widetilde{\Psi}(T_1)-\widetilde{\Psi}(T_2)$,
which shows $\widehat{\Psi}$ is well defined. Furthermore, for
$T\in{\mathcal B}_{1r}(H)$, it is easy to see
$\widehat{\Psi}(-T)=-\widehat{\Psi}(T)$, combining the homogeneity
of $\widetilde{\Psi}$ over non-negative real number, we know
$\widehat{\Psi}$ is linear. Assume $\widehat{\Psi}(T)=0$, from the
definition of $\widehat{\Psi}$,
$\widehat{\Psi}(T^+)=\widehat{\Psi}(T^-)$, i.e.,
$\widetilde{\Psi}(T^+)=\widetilde{\Psi}(T^-)$. Now, the
injectivity of $\widetilde{\Psi}$ implies $T^+=T^-$, so $T=0$ and
$\widehat{\Psi}$ is injective.
 If
$\Gamma:{\mathcal B}_{1r}(H)\rightarrow{\mathcal B}_{1r}(H)$ is
another  linear map which extends $\Psi$, then for any
$T\in{\mathcal B}_{1r}(H)$, $$\begin{aligned}
\Gamma(T)&=\Gamma(T^+-T^-)=\Gamma(T^+)-\Gamma(T^-)\\
       &=\|T^+\|_1\Gamma(\frac{T^+}{\|T^+\|_1})-\|T^-\|_1\Gamma(\frac{T^-}{\|T^-\|_1})\\
       &=\|T^+\|_1\Psi(\frac{T^+}{\|T^+\|_1})-\|T^-\|_1\Psi(\frac{T^-}{\|T^-\|_1})\\
       &=\widehat{\Psi}(T^+)-{\widehat\Psi}(T^-)={\widehat\Psi}(T)
       .\end{aligned}$$
This shows the extension is unique, as desired.

Now, ${\widehat\Psi}:{\mathcal B}_{1r}(H)\rightarrow{\mathcal
B}_{1r}(H)$ is a linear injection. We assert that ${\widehat\Psi}$
is continuous in the trace norm $\|.\|_1$. For any $T\in S_1(H)$,
clearly, $\|{\widehat\Psi}(T)\|_1=\|\Phi(T)-\Phi(0)\|_1\leq 2$.
For arbitrary $T\in {\mathcal B}_{1r}(H),\|T\|_1\leq 1$, it is
easy to see $T^+\in S_1(H), T^-\in S_1(H)$. Thus
$\|{\widehat\Psi}(T)\|_1=\|{\widehat\Psi}(T^+)-{\widehat\Psi}(T^-)\|_1
\leq\|{\widehat\Psi}(T^+)\|_1+\|{\widehat\Psi}(T^-)\|_1\leq 4$. It
follows that ${\widehat\Psi}$ is bounded on the unit ball of
${\mathcal B}_{1r}(H)$, hence ${\widehat\Psi}$ is continuous. Note
that $\Psi$ is the restriction of ${\widehat\Psi}$ on $S_1(H)$,
therefore $\Psi$ is continuous and so $\Phi$ is continuous on
$S_1(H)$, as desired.

{\bf Claim 2.} $\Phi(0)=0$,\hspace{0.1in}$\Phi({\mathcal
P}_1(H))={\mathcal P}_1(H)$.

Clearly, $\Phi$  preserves the extreme points of $S_1(H)$ which
are exactly the one dimensional projections and zero in $S_1(H)$
(see \cite[Lemma 1]{BD}). Since $\Phi^{-1}$ has the same
properties as $\Phi$, $\Phi({\mathcal P}_1(H)\cup\{0\})={\mathcal
P}_1(H)\cup\{0\}$. We claim that $\Phi(0)=0$. Assume on the
contrary, $\Phi(0)\neq 0$, then there exists one dimensional
projection $P$ such that $\Phi(P)=0$. Note that for $P_1, P_2\in
{\mathcal P}_1(H)$, $\|P_1-P_2\|_1=2\sqrt{1-tr(P_1P_2)}$.
Therefore we can choose a sequence  $\{P_n\}_{n=1}^{\infty}$ in
${\mathcal P}_1(H)$ such that $\|P_n-P\|_1\rightarrow
0(n\rightarrow +\infty)$. By Claim 1 and the property of $\Phi$ ,
 $\|\Phi(P_n)-\Phi(P)\|_1=1\rightarrow 0(n\rightarrow
+\infty)$, a contradiction. This tells us $\Phi(0)=0$ and so
$\Phi({\mathcal P}_1(H))={\mathcal P}_1(H)$.

{\bf Claim 3.} There exists an either unitary or antiunitary
operator $U$ on $H$ such that $\Phi(T)=UTU^*$ for all $T\in
S_1(H)$.

From Claim 2, $\Phi=\Psi$, so we denote
$\widehat{\Psi}=\widehat{\Phi}$. From the proof of Claim 1, one
can easily get $\widehat{\Phi}$ is also surjective. Thus $\Phi$
has a unique positive linear bijective extension from $S_1(H)$ to
${\mathcal B}_{1r}(H)$. In addition, Since $\Phi^{-1}$ has the
same properties as $\Phi$, a direct computation shows
$\widehat{\Phi^{-1}}={\widehat{\Phi}}^{-1}$.

In the following, we will prove ${\widehat\Phi}$ is trace norm
preserving. Firstly, it will be shown that ${\widehat\Phi}$ is
trace preserving, i.e. $tr(T)=tr({\widehat\Phi}(T))$ for every
$T\in{\mathcal B}_{1r}(H)$. Assume $T\in{\mathcal B}_{1r}^+(H)$,
and $T=\lambda_1 P_1+\lambda_2 P_2+\cdots+\lambda_nP_n$, where
$\{P_i\}_{i=1}^n$ are mutually orthogonal one dimensional
projections, $\lambda_h>0, i=1,2\cdots,n$. Then
$$\begin{aligned}
{\widehat\Phi}(T)&=\lambda_1{\widehat\Phi} (P_1)+\lambda_2
{\widehat\Phi}(P_2)+\cdots+\lambda_n{\widehat\Phi}(P_n)\\
                 &=\lambda_1\Phi(P_1)+\lambda_2 \Phi(P_2)+\cdots+\lambda_n\Phi(P_n).\end{aligned}$$
By Claim 2,  we can obtain
$tr(T)=\Sigma_{i=1}^n\lambda_i=tr({\widehat\Phi}(T))$. For any
$T\in{\mathcal B}_{1r}^+(H)$, by the spectral theorem of positive
operators, there exists monotone increasing sequence
$\{T_n=\Sigma_{i=1}^n\lambda_iP_i\}_{n=1}^\infty$ such that
$\|T_n-T\|_1=tr(T-T_n)=tr(T)-tr(T_n)\rightarrow 0(n\rightarrow
\infty)$, where $\{P_i\}_{i=1}^n$ are mutually orthogonal one
dimensional projections, $\lambda_i>0, i=1,2,\cdots,n$. Since
${\widehat\Phi}$ is positive preserving and continuous,
$\{{\widehat\Phi}(T_n)\}_{n=1}^\infty$ is monotone increasing and
$\|{\widehat\Phi}(T_n)-{\widehat\Phi}(T)\|_1=tr({\widehat\Phi}(T))-tr({\widehat\Phi}(T_n))\rightarrow
0(n\rightarrow\infty)$. Note that
$tr({\widehat\Phi}(T_n))=tr(T_n)$, so for every $T\in{\mathcal
B}_{1r}^+(H), tr(T)=tr({\widehat\Phi}(T))$. For any $T\in{\mathcal
B}_{1r}(H)$,
$$tr({\widehat\Phi}(T))=tr({\widehat\Phi}(T^+))-tr({\widehat\Phi}(T^-))=tr(T^+)-tr(T^-)=tr(T),$$
So ${\widehat\Phi}:{\mathcal B}_{1r}(H)\rightarrow{\mathcal
B}_{1r}(H)$ is positive and  trace preserving.

Next, we will show that ${\widehat\Phi}$ preserves the trace norm.
In fact, for any $T\in{\mathcal B}_{1r}(H)$, we have
$$\begin{aligned}
\|{\widehat\Phi}(T)\|_1 &= \|{\widehat\Phi}(T^+-T^-)\|_1=\|{\widehat\Phi}(T^+)-{\widehat\Phi}(T^-)\|_1         \\
                        &\leq \|{\widehat\Phi}(T^+)\|_1+\|{\widehat\Phi}(T^-)\|_1=tr({\widehat\Phi}(T^+))+tr({\widehat\Phi}(T^-))\\
                        &=tr(T^+)+tr(T^-)=tr(T^++T^-)=tr(|T|)=\|T\|_1.
\end{aligned}$$
So ${\widehat\Phi}:{\mathcal B}_{1r}(H)\rightarrow{\mathcal
B}_{1r}(H)$ is contractive, i.e., for $T\in{\mathcal
B}_{1r}(H),\|{\widehat\Phi}(T)\|_1\leq\|T\|_1$. Since
${\widehat\Phi}^{-1}$ has the same properties as ${\widehat\Phi}$,
we have $\|{\widehat\Phi}(T)\|_1\geq\|T\|_1$ and thus
$\|{\widehat\Phi}(T)\|_1=\|T\|_1$, that is ${\widehat\Phi}$ is a
$\|\cdot\|_1$-isometry of ${\mathcal B}_{1r}(H)$.

Note that, for $P,Q\in{\mathcal P}_1(H),
\|P-Q\|_1=2\sqrt{1-tr(PQ)}$. Thus $PQ=0\Leftrightarrow
\|P-Q\|_1=2$. Since ${\widehat\Phi}$ is trace norm preserving, we
have $PQ=0\Leftrightarrow{\widehat\Phi}(P){\widehat\Phi}(Q)=0$. By
Claim 2, ${\widehat\Phi}|_{{\mathcal P}_1(H)}:{\mathcal
P}_1(H)\rightarrow{\mathcal P}_1(H)$ is a bijection with the
property
$PQ=0\Leftrightarrow{\widehat\Phi}(P){\widehat\Phi}(Q)=0,P,Q\in{\mathcal
P}_1(H)$. Using the well-known Uhlhorn-type of Wigner's theorem
(see \cite{Uh}), we have ${\widehat\Phi}(P)=UPU^*(P\in{\mathcal
P}_1(H))$ for some unitary or antiunitary operator $U$ on $H$. By
the spectral theorem of selfadjoint operators  and the continuity
of ${\widehat\Phi}$, for all $T\in{\mathcal B}_{1r}(H)$,
${\widehat\Phi}(T)=UTU^*$, therefore $\Phi(T)=UTU^*$ for all $T\in
S_1(H)$, as desired.

Based on the Theorem 2.1, we can prove Corollary 2.2.

{\bf Proof of Corollary 2.2.}  Firstly, we recall a nice result of
Mankiewicz, namely \cite[Theorem 5]{Man} which states that if we
have a bijective isometry between convex set in normed linear
space with nonempty interiors, then this isometry can be uniquely
extended to a bijective affine isometry between the whole space.
Clearly, in the finite dimensional case, the convex set $S_1(H)$
has  nonempty interiors in the normed linear space of ${\mathcal
B_{1r}}(H)$ (In fact, the interior of $S_1(H)$ consists of all
invertible positive operators). Consequently, applying the result
of Mankiewicz, we know that $\Phi$ is automatically affine.
Combing this with Theorem 1, we get the desired.

\textbf{Remark 2.3.} Now, in order to illustrate that the
bijective assumption is indispensable in theorem 2.1 and corollary
2,2. we give an example, the idea is come from the Kraus
representation theorem (see \cite{ Kra}): Suppose that $H$ is a
complex separable infinite dimensional Hilbert space such that $H$
can be presented as a direct sum of mutually orthogonal closed
subspaces, $H=(\oplus_{k=1}^N H_k)\oplus H_0$, $N\in{\mathbb N}$,
$\dim H_k=\dim H$, $k=1,2,\cdots, N$. Let $U_k:H\rightarrow H_k$
be unitary or antiunitary operators, $\lambda_1, \lambda_2,
\cdots, \lambda_{N}\in(0,1), \sum_{k=1}^{N}{\lambda_k}=1$. Let
$\Phi(T)=\sum_{k=1}^{N} \lambda_kU_kTU_k^*,\hspace{0.1in} \forall
T\in S_1(H)$. Then $tr(\Phi(T))=\sum_{k=1}^{N}
\lambda_ktr(U_kTU_k^*)=tr(T)$, the last equality being due to
$U_k^*U_k=I$. This implies that $\Phi$ is indeed a mapping which
maps $S_1(H)$ into $S_1(H)$. Furthermore, it is easy to see that
$\Phi(\lambda T+(1-\lambda)S)=\lambda\Phi(T)+(1-\lambda)\Phi(S)
\hspace{0.1in}\mbox{for all} \hspace{0.1in}T,S\in
S_1(H),0\leq\lambda\leq 1$. Finally, all $\Phi_k:T\rightarrow
U_kTU_k^*$ are isometric and for all $T\in S_1(H),
|\Phi_k(T)|\bot|\Phi_l(T)|(\mbox{i.e.}, |\Phi_k(T)||\Phi_l(T)|=0)$
if $k\neq l$. Thus
$$\|\Phi(T)-\Phi(S)\|_1=\|\sum_{k=1}^{N}
\lambda_kU_k(T-S)U_k^*\|_1=\sum_{k=1}^{N}
\lambda_k(\|T-S\|_1)=\|T-S\|_1$$ for all $T,S\in S_1(H)$. But, in
general, such $\Phi$ is not a bijection and does not have a nice
form as theorem 2.1 and corollary 2.2.

\section{ Ortho-order
automorphisms on $S_1(H)$ }

The purpose of this section is to characterize the ortho-order
automorphisms of $S_1(H)$, that is, the bijective map $\Phi$
preserves the spectral order in both directions and preserves
orthogonality in both directions. The following is the main
result.

 \textbf{Theorem 3.1.} {\it  If $\Phi:S_1(H)\rightarrow S_1(H)$ is
an ortho-order automorphism, then there exists an either unitary
or antiunitary operator $U$ on $H$, a strictly increasing
continuous bijection $f:[0,1]\longmapsto[0,1]$ such that
$\Phi(T)=Uf(T)U^*$ for all $T\in S_1(H)$, where $f(T)$ is obtained
from the continuous function calculus.}

Before embarking on the proof of Theorem 3.1, we need some
terminologies and facts about spectral order.

First of all, by a resolution of identity we mean a function from
${\mathbb R}$ into the lattice $({\mathcal P}(H),\leq)$ of all
projections on $H$ which is increasing, right-continuous, for all
small real numbers it takes the value $0$, while for large enough
real numbers it takes value $I$ (the identity operator of $H$). It
is well-known that there is a one-to-one correspondence between
the compactly supported spectral measures on the Borel sets of
${\mathbb R}$ and the resolutions of the identity (see \cite[Page
360]{Kad2}). In fact, every resolution of the identity is the form
$t\mapsto E(-\infty,t]$. If $T\in S_1(H)$, the resolution of the
identity corresponding to $E_T$ is called the spectral resolution
of $T$. Next, the spectral order implies the usual order: if
$T,S\in S_1(H)$ and $T\preceq S$, then $T\leq S$. Furthermore,
$T\preceq S$ if and only if $T^n\leq S^n$ for every $n\in{\mathbb
N}$. For commuting $T,S\in S_1(H)$, by the spectral theorem of
positive operators, it is easy to see $T\preceq S$ if and only if
$T\leq S$. Finally, for $T,S\in S_1(H)$, the supremum of the set
$\{T,S\}$ in this structure denoted by $T\vee S$  exists.
Similarly,  the infimum of the set $\{T,S\}$ denoted by $T\wedge
S$ also exists. For details, one can see \cite{Ols}.

After these preparations, we turn to the proof of Theorem 3.1.

{\bf Proof of Theorem 3.1.} The proof is divided into 3 claims.

{\bf Claim 1.} $\Phi$ preserves the rank of operators.

Let $\Phi$ be an ortho-order automorphism of $S_1(H)$. Since
$0=\wedge S_1(H)$, it follows that $\Phi(0)=0$. For $T\in S_1(H)$,
we denote by $\{T\}^\bot=\{S\in S_1(H):TS=0\}$, i.e., the set of
all elements of $S_1(H)$ which are orthogonal to $T$. By the
spectral theorem of positive operators, it is easy to see that
$T\in S_1(H)$ is of rank $n$ if and only $\{T\}^{\bot\bot}$
contains $n$ pairwise orthogonal nonzero elements but it does not
contain more. As $\Phi$ preserves orthogonality in both
directions, it is now clear that $\Phi$ preserves the rank of
operators.

{\bf Claim 2.} There exists a strictly increasing continuous
bijection $f:[0,1]\longmapsto[0,1]$ such that $\Phi(\lambda
P)=f(\lambda )\Phi(P)$ for all $P\in {\mathcal P}_1(H)$.

By Claim 1, $\Phi$ preserves the rank of operators. In particular,
$\Phi$ preserves the rank one elements of $S_1(H)$. Since
$\Phi^{-1}$ has the same properties as $\Phi$, we have $\Phi$
preserves the rank one elements in both directions, i.e., $T\in
S_1(H)$ is rank one if and only if $\Phi(T)$ is rank one. Note
that the rank one projections are exactly the maximal elements of
the set of all rank one elements in $S_1(H)$. This implies $\Phi$
preserves rank one projections in both directions, i.e.,
$\Phi({\mathcal P}_1(H))={\mathcal P}_1(H)$.

Let $P$ be a rank one projection. For $\lambda\in[0,1]$, then
$\lambda P\preceq P$ and so we have $\Phi(\lambda
P)\preceq\Phi(P)$. This implies that there is a scalar
$f_P(\lambda)\in[0,1]$ such that $$\Phi(\lambda
P)=f_P(\lambda)\Phi(P).$$ It follows from the properties of $\Phi$
 that $f_P$ is a strictly increasing continuous
bijection of $[0,1]$. Now, $\Phi(0)=0$ together with $\Phi$
preserves rank one projection implies $f_P(0)=0,f_P(1)=1$.

In the following, we will prove that $f_P$ does not depend on $P$.

Let $E,F,E\neq F$, be rank one projections and
$0<\lambda\leq\mu\leq 1$. Computing the spectral resolution of
$\lambda E$ and $ \mu F$, we have $$E_{\lambda
E}(-\infty,t]=\left\{\begin{array}{cc} 0 & t<0;\\
                                       I-E & 0\leq t<\lambda;\\
                                       I & \lambda\leq t,
                                       \end{array}\right.$$
$$E_{\mu
F}(-\infty,t]=\left\{\begin{array}{cc} 0 & t<0;\\
                                       I-F & 0\leq t<\mu;\\
                                       I & \mu\leq t.
                                       \end{array}\right.$$
From \cite{Ols}, we know that $E_{\lambda E\vee \mu
F}(-\infty,t]=E_{\lambda E}(-\infty,t]\wedge E_{\mu
F}(-\infty,t]$, and so $$E_{\lambda E\vee
\mu F}=\left\{\begin{array}{cc} 0 & t<0;\\
                                       (I-E)\wedge(I-F) & 0\leq t<\lambda;\\
                                       I-F &\lambda\leq t<\mu;\\
                                       I & \mu\leq t.
 \end{array}\right.$$
Note that $(I-E)\wedge(I-F)=I-E\vee F$, thus we have  $$\lambda
E\vee \mu F=\lambda(E\vee F-F)+\mu F.$$ This tells us that the
nonzero eigenvalues of the operator $\lambda E\vee \mu F$ are
$\lambda$ and $\mu$.

Let $R$ be a rank two projection of $H$ and pick
$\lambda\leq\frac{1}{2}$, then $\lambda R\in S_1(H)$. Since $\Phi$
preserves the rank of operators, we have $\Phi(\lambda R)$ is a
rank two operator and hence it can be written in the form
$$\Phi(\lambda R)=\alpha P'+\beta Q'$$ with mutually orthogonal
rank one projection $P',Q'$ and $0<\alpha\leq\beta<1$. Pick any
two different rank one subprojections $P,Q$ of $R$. Then we
compute
$$\Phi(\lambda R)=\Phi(\lambda P\vee\lambda Q)=\Phi(\lambda
P)\vee\Phi(\lambda Q)=f_P(\lambda)\Phi(P)\vee
f_Q(\lambda)\Phi(Q).$$ It follows that
$$\{\alpha, \beta\}=\{f_P(\lambda),f_Q(\lambda)\}.$$

We claim that $\alpha=\beta$. Suppose on the contrary that
$\alpha\neq\beta$. Without loss of generality, we may assume that
$f_P(\lambda)=\alpha$ and $f_Q(\lambda)=\beta$. Pick a third rank
one subprojection $R_0$ of $R$ which is different from $P$ and
$Q$. Then repeating the above argument for the pair $P,R_0$, we
have $f_{R_0}(\lambda)=\beta$. Similarly, for the pair $R_0,Q$, we
have $f_{R_0}(\lambda)=\alpha$. This contradiction yields that
$\alpha=\beta$ and so $f_P(\lambda)=f_Q(\lambda)$ for all
$\lambda\in[0,\frac{1}{2}]$. Thus we can denote
$f_P(\lambda)=f(\lambda)$ for all $\lambda\in[0,\frac{1}{2}]$.

The remianed is to prove $f_P(\lambda)$ does not depend on $P$ for
every $\lambda\in(\frac{1}{2},1)$. Let $Q$ be a rank one
projection such that $PQ=0$. We compute $$\Phi(\lambda
P+(1-\lambda)Q)=\Phi(\lambda P\vee(1-\lambda) Q))=\Phi(\lambda
P)\vee\Phi((1-\lambda) Q).$$  Since $\Phi(\lambda P)$ and
$\Phi((1-\lambda)Q)$ are orthogonal, It follows  that
$$\Phi(\lambda P+(1-\lambda)Q)=f_P(\lambda)\Phi(
P)+f(1-\lambda)\Phi(Q).$$ For $T\in S_1(H)$, by the spectral
theorem of positive operator,  $tr(T)=1$ if and only if there does
not exist $S\in S_1(H)$ such that $T\preceq S$. By the properties
of $\Phi$ and $\Phi^{-1}$, we have $\Phi(V)=V$, recall that $V$ is
the surface of $S_1(H)$. Combing this with $\Phi(\lambda
P+(1-\lambda)Q)=f_P(\lambda)\Phi(P)+f(1-\lambda)\Phi(Q)$, we can
obtain $f_P(\lambda)+f(1-\lambda)=1$. Clearly,
$f_P(\lambda)=1-f(1-\lambda)$ and so $f_P(\lambda)$ does not
depend on $P$ for every $\lambda\in(\frac{1}{2},1)$. This
completes the proof of this claim.

{\bf Claim 3.} There exists an either unitary or antiunitary
operator $U$ on $H$, a strictly increasing continuous bijection
$f:[0,1]\longmapsto[0,1]$ such that $\Phi(T)=Uf(T)U^*$ for all
$T\in S_1(H)$, where $f(T)$ is obtained from the continuous
function calculus.

Now $\Phi:{\mathcal P}_1(H)\rightarrow{\mathcal P}_1(H)$ is a
bijection and preserves orthogonality in both directions. By the
Uhlhorn-type of Wigner's theorem (see \cite{Uh}), there exists a
unitary or antiunitary operator $U$ on $H$ such that
$\Phi(P)=UPU^*$ for all $P\in{\mathcal P}_1(H)$.

For $T_n=\lambda_1P_1+\lambda_1P_2+\cdots+\lambda_nP_n$,
$\lambda_i\in(0,1](i=1,2,\cdots,n),\sum_{i=1}^{n}\lambda_i\in(0,1]$,
$P_iP_j=0(i\neq j, i,j=1,2,\cdots n)$. Then
$$\begin{aligned}\Phi(T)&=\Phi(\lambda_1P_1+\lambda_1P_2+\cdots+\lambda_nP_n)\\
                        &=\Phi(\lambda_1P_1\vee\lambda_2P_2\vee\cdots\vee\lambda_nP_n)\\
                        &=f(\lambda_1)UP_1U^*+\cdots+f(\lambda_n)UP_nU^*\\
                        &=Uf(T)U^*,\end{aligned}$$where $f(T)$ is obtained from the continuous
function calculus. For every $T\in S_1(H)$, by the spectral
theorem of positive operators, there exists a monotonically
increasing sequence $\{T_n\}_{n=1}^{+\infty}$ of $S_1(H)$ such
that $T=\vee_{n=1}^{+\infty} T_n$. Since $\Phi$ preserves the
spectral order of operators in both directions, it follows that
$\Phi(T)=Uf(T)U^*$ for every $T\in S_1(H)$.

{\bf Acknowledgments}

This work was supported partially by National Natural Science
Foundation of China (10771175,11071201),Youth National Natural
Science Foundation of China (11001230) , the Fundamental Research
Funds for the Central Universities(2010121001) and Youth Talented
Natural Science Foundation  of Fujian (2008F3103)

\end{document}